\documentstyle[aps,preprint,epsfig]{revtex}
\newcommand{\beq}{\begin{equation}}
\newcommand{\eeq}{\end{equation}}
\newcommand{\bef}{\begin{figure}}
\newcommand{\enf}{\end{figure}}
\newcommand{\bdis}{\begin{displaymath}}
\newcommand{\edis}{\end{displaymath}}

\title{Drifter dispersion in the Adriatic Sea: Lagrangian data 
and chaotic model 
}
\author{Guglielmo Lacorata$^1$, Erik Aurell$^2$, Angelo Vulpiani$^3$}
\begin{document}
\maketitle

\centerline{$^1$ Dipartimento di Fisica, Universit\`{a} dell'Aquila}
\centerline{Via Vetoio 1, I-67010 Coppito, L'Aquila, Italy, and}
\centerline{Istituto di Fisica dell'Atmosfera, CNR}
\centerline{Via Fosso del Cavaliere, I-00133 Roma, Italy}
\centerline{$^2$ Department of Mathematics, Stockholm University}
\centerline{S-106 91 Stockholm, Sweden}
\centerline{$^3$ Istituto Nazionale Fisica della Materia, 
Unit\`{a} di Roma 1, and}
\centerline{Dipartimento di Fisica, Universit\`{a} di Roma 
"La Sapienza"}
\centerline{Piazzale Aldo Moro 5, I-00185 Roma, Italy}

\medskip

\date{\today}

\begin{abstract}

\noindent {We analyze characteristics of drifter trajectories  
from the Adriatic Sea with recently introduced
nonlinear dynamics techniques.
We discuss how in quasi-enclosed basins,  
relative dispersion as function of time,
a standard analysis tool in this context,
may give a distorted picture of the dynamics.
We further show that useful information may
be obtained by using two related
non-asymptotic   
indicators, the Finite-Scale Lyapunov Exponent (FSLE) and
the Lagrangian Structure Function (LSF),
which both describe intrinsic physical properties 
at a given scale. We introduce a simple chaotic model for 
drifter motion in this system,  
and show by comparison with the model
that Lagrangian dispersion is mainly 
driven by advection at sub-basin scales until saturation sets in. 
}

\end{abstract}



\newpage

\renewcommand{\baselinestretch}{2.} 

\section{Introduction}

Understanding the mechanisms of 
transport and mixing processes is an important 
and challenging task, which has wide relevance from a theoretical 
point of view, e.g. for the study of diffusion and chaos in geophysical 
systems in general, or for validating 
simulation results from a general circulation model. 
It is also a necessary tool in the analysis 
of problems of general interest and social impact, such as 
the dispersion of nutrients or
pollutants in sea water with consequent 
effects on marine life and on the environment 
(Adler et al., 1996). 

Recently, a number of oceanographic programs have been devoted to the 
study of the surface circulation of the Adriatic Sea by  
the observation of Lagrangian drifters, 
within the larger framework 
of drifter-related research in the whole Mediterranean Sea 
(Poulain, 1999).  

The Adriatic Sea is a quasi-enclosed basin, about 800 long by 200 $km$ wide, 
connected to the rest of the Mediterranean Sea through the Otranto Strait. 
From a topographic point of view, three major regions can be considered:
the northern part is the shallowest, about 100 $m$ maximum depth,  
and extends down to the latitude of Ancona;  
the central part, which deepens down to 
about 260 $m$ in the Jabuka Pit,
and the southern part which extends from the Gargano promontory to the Otranto 
Strait. The southern part is the deepest, 
reaching about 1200 $m$ in the South Adriatic Pit. 
Reviews on the oceanography of the Adriatic Sea can be found in 
Artegiani et al. (1997), Orlic et al. (1992), Poulain (1999) and Zore (1956).  

Lagrangian data offer the opportunity  
to employ techniques of analysis, well established in the theory 
of chaotic dynamical systems, to study the behavior of  
actual trajectories and compare those with a 
kinematic model.  

Let us assume that the 
Lagrangian drifters are passively advected in a 
two-dimensional flow, e.g. as would
be the case in a frictionless barotropic approximation   
(Ottino, 1989; Crisanti et al., 1991):
\beq
{dx \over dt} \,=\, u(x,y,t) \;\; \hbox{and} \;\; 
{dy \over dt} \,=\, v(x,y,t) \, ,
\eeq
where $(x(t), y(t))$ is the position of 
a fluid particle at time $t$ in terms of longitude and latitude and 
$u$ and $v$ are the zonal and meridional 
velocity fields, respectively. 

For the Eulerian description of a geophysical system, 
one should in principle use
numerical solutions of the Navier-Stokes equations 
(or other suitable equations, e.g. the quasi-geostrophic model) 
to obtain the velocity fields. 
In practice, direct numerical simulation of these equations
on oceanographic length scales is of course not possible,
and one has to invoke approximations, i.e. turbulence modeling.
This motivates to use instead a simplified kinematic 
approach, by adopting a given Eulerian velocity field. 
The criteria for the construction of such a field follow from phenomenological 
arguments and/or experimental observation, and have recently
been reviewed in
(Yang, 1996;  Samelson, 1996). 

Let us consider the relationship between 
Eulerian and Lagrangian properties of a system. A wide literature 
on this topic (e.g. Ottino, 1989; Crisanti et al., 1991)  
allows us state that, 
in general, 
motion in Eulerian and Lagrangian variables can be rather 
different. 
It is not rare to have regular Eulerian behavior, 
e.g. a time-periodic velocity field, co-existing with Lagrangian 
chaos or vice-versa. 

In quasi-enclosed basins like the 
Adriatic Sea, a characterization of the mechanisms 
of the mixing is highly non-trivial. We first observe,
see below for detailed discussion,
that the use of 
the standard diffusion coefficients can have rather limitated
applicability 
(see Artale et al., 1997). 
Already classical studies on
Lagrangian 
particles in ocean models contain remarks on 
the intrinsic difficulties in using one-particle diffusion statistics  
(Taylor, 1921). In situations where the advective 
time is not much longer than 
the typical decorrelation time scale of the Lagrangian velocity, the 
diffusivity parameter related to small-scale turbulent motion cannot 
converge to its asymptotic value (Figueroa and Olson, 1994). 
One the other hand, a generalization of the standard 
Lyapunov exponent, the Finite-Scale Lyapunov 
Exponent (FSLE), originally introduced for the predictability problem
(Aurell et al., 1996:1997), has been shown to be 
a suitable tool to describe non-asymptotic 
properties of transport. This finite-scale approach to 
Lagrangian transport 
measures effective rates of particle dispersion
without assumptions 
about small-scale turbulent processes.  
For an alternative 
method see Buffoni et al. (1997);
for a recent review and systematic discussion of
non-asymptotic properties of transport and mixing in realistic cases, see 
Boffetta et al. (2000).

In this paper we report data analysis
of  surface drifter motion in the 
Adriatic Sea using  
relative dispersion, FSLE
and Lagrangian Structure Function (LSF),
a quantity related to the FSLE. 
We also introduce a chaotic model for the Lagrangian 
dynamics, and use the FSLE and LSF characteristics to compare
model and data.
We show    
that it can be very difficult to get an
estimate of the diffusion coefficient in
a quasi-enclosed basin, and/or 
to look for deviations from the 
standard diffusion law.
In fact, the time a cluster 
of particles takes to spread uniformly and reach the boundaries 
is not much longer than the largest characteristic 
time of the system.  
In contrast, the FSLE and the LSF  
do characterize the transport 
properties of Lagrangian trajectories at a fixed spatial scale. 
We will finally show that a simple kinematic model
reproduces the data.

In section 2 we describe   
the data set we have used, review relevant 
concepts and analysis 
techniques for 
Lagrangian transport and chaos. 
In section 3 we introduce a kinetic model of 
the Lagrangian dynamics, and in section 
4 we compare the data and the model.  
Section 5 contains a summary and a discussion  
of the results.

\section{Data set and analysis techniques}

\subsection{Data set}

In a large drifter research program
in 
Mediterranean Sea, started in the late 80's and continued into the 90's, 
Lagrangian data 
from surface drifters  
deployed in the Adriatic sea 
have been recorded 
from December 1994 to March 1996. 
These drifters are 
similar to the CODE (COastal Dynamics Experiment) system (Davis, 1985) 
and they are designed to 
be sufficiently wind-resistant so to effectively give a description of 
the circulation at their actual depth (1 meter).  
The drifters were tracked by the Argos Data Location and Collection 
System (DCLS) carried by the NOAA polar-orbiting satellites. It is assumed 
that after data processing drifter positions are accurate to within 
200-300 $m$, and velocities to within
2-3 $cm/s$. 

For a description of the experimental program, see Poulain (1999). Technical 
details about the treatment of raw data can also be found in 
Hansen and Poulain (1996), Poulain et al. (1996) and
Poulain and Zanasca (1998).  

The data have been stored in separate files, one for  
each drifter. In the format used by us, each file contains: 
number of records (i.e. number of points of the trajectory); 
time in days; position of the drifter in longitude and latitude; 
velocity of the drifter along the zonal and meridional directions;
and 
temperature in centigrade degrees. The sampling time is 6 hours. 
We can identify five main deployments on which 
we will concentrate our attention. Selecting the tracks by the time of 
the first record, it is easy to verify that these five subsets 
consist of drifters 
deployed in the same area in the Otranto
strait, near 19 degrees longitude east and 
40 degrees latitude north.
The experimental strategy of simultaneously
releasing the drifters 
within a distance of some 
kilometers, allows us to study
dispersion quantitatively.

From a qualitative point of view, what we observe from 
the plot of all the trajectories (Fig. 1) 
is the shape of two (cyclonic) 
basin-wide gyres, located in the middle and southern regions 
respectively, 
and, 
 an anti-clockwise boundary current which moves the drifters 
north-westward along the east coast and south-eastward down the west coast.
The latter is a permanent feature of the Adriatic sea
(Poulain, 1999). 
On the other hand,
it is known that, within a year, the pattern of basin-wide
gyres may change between one, two and three
gyres over a time-scale of months.
The southern gyre is the 
most steady of the three.  
The data also suggest the presence of 
small scale structures, even though these are likely much more variable 
in time. 
The time-scale of the typical recirculation period around a basin-wide gyre 
is about one month and the time needed to travel along 
the coasts and complete one lap of the full basin is of the order of 
a few months. 

\subsection{Analysis techniques}

We recall here some basic concepts about dynamical systems, 
diffusion and chaos, and the quantities that we shall use to characterize 
the properties of Lagrangian trajectories.

If we have $N_c$ clusters of initially close particles, 
each cluster containing $n_k$ elements, 
relative dispersion can be characterized by the diffusion
coefficient 
\beq
D_i \,=\, {\hbox {lim}}_{t \to \infty} \, 
{1 \over 2t} \, S^{2}_{i}(t)
\eeq
with 
\beq
S^{2}_{i}(t) \,=\,
{1 \over N_c} \sum_{k=1}^{N_c}{1 \over n_k} 
\sum_{j=1}^{n_k} (x_i^{(k,j)}(t)-<x_i(t)>^{(k)})^2
\eeq
where 
\beq
<x_i(t)>^{(k)} \,= \, {1 \over n_k} \sum_{j=1}^{n_k}x_i^{(k,j)}(t)
\eeq
$x_i^{(k,j)}$ is the $i$-th spatial coordinate of the $j$-th particle 
in the $k$-th cluster; $S^{2}=\sum_iS^{2}_{i}$ 
is  
the mean square displacement of the particles relatively to their 
time evolving mean position. 
If $\delta(t)$ is the distance between two trajectories 
${\bf x}^{(1)}$ and ${\bf x}^{(2)}$ in a cluster 
at time $t$, relative dispersion is defined as
\beq
<\delta^2(t)> \,=\, <||{\bf x}^{(1)}(t)-{\bf x}^{(2)}(t)||^2>
\eeq
where the average is over all pairs of
trajectories in the cluster.   
In a standard diffusive regime, ${\bf x}^{(1)}(t)$ and ${\bf x}^{(2)}(t)$ 
become independent variables and,  
 for $t \to \infty$, 
we have $<\delta^2(t)> = 2 \cdot S^{2}(t)$.
In the following, 
we shall consider the cluster mean square radius $S^{2}(t)$ as 
a measure of relative dispersion. Absolute dispersion, which 
is defined as the mean square displacement from an initial position  
will not be taken into account in our analysis. 

If, in the asymptotic limit, 
$S^{2}_{i}(t) \sim t^{2\alpha}$ with $\alpha=1/2$, we have the 
linear law of standard 
diffusion 
for the mean square displacement, and the $D_i$'s are finite; 
if $\alpha \neq 1/2$ we have so-called anomalous diffusion 
(Bouchaud and Georges, 1990). 

The difficulty that often arises when measuring the exponent $\alpha$ is 
that, because of the finite size of the domain,  
dispersion cannot reach its true asymptotic behavior. In other words, 
diffusion may not be observable over sufficiently large scales, i.e. much 
larger than the largest Eulerian length scale, 
and, therefore, we cannot have a robust estimate of the exponent of the 
asymptotic power law. 
Moreover, the relevance of asymptotic quantities, 
like the diffusion coefficients, is questionable 
in the study of realistic cases 
concerning the transport problem in finite-size systems (Artale et al., 1997).

The diffusion coefficients characterize long-time (large-scale) dispersion 
properties. In contrast, at short times (small scales) the relative 
dispersion is related to the chaotic behavior of the Lagrangian trajectories.

A quantitative measure of instability for the time evolution 
of a dynamical system (Lichtenberg and Lieberman, 1992) 
is commonly given by the Maximum Lyapunov Exponent (MLE) 
$\lambda$, which gives the rate of exponential separation of two nearby 
trajectories
\beq
\lambda \,=\, {\hbox{lim}}_{t \to \infty} \, {\hbox{lim}}_{\delta(0) \to 0} \, 
{1 \over t} \, \ln \, {\delta(t) \over \delta(0)}
\eeq
where $\delta(t)=||{\bf x}^{(1)}(t)-{\bf x}^{(2)}(t)||$ 
is the distance between 
two trajectories at time $t$.
When $\lambda > 0$ the system is said to be chaotic. 
There exists a well established algorithm to numerically compute the 
MLE introduced by Benettin et al. (1980).

A characteristic time, $T_{\lambda}$, associated to the MLE is the 
predictability time, defined as the minimum time 
after which the error on the state of the system becomes larger than a 
tolerance value $\Delta$, if the initial uncertainty is $\delta$ 
(Lichtenberg and Lieberman, 1992): 
\beq
T_{\lambda} \,=\, {1 \over \lambda} \, \ln \, {\Delta \over \delta}
\eeq

Let us recall that $\lambda$ is a mathematically well-defined quantity 
which measures the growth of infinitesimal errors.  
In physical terms, at any time, $\delta$ has to be much lesser than the 
characteristic size of the smallest relevant length of the velocity field. 
For example, in 3-D fully developed turbulence $\delta$ has to be 
much smaller than the Kolmogorov length. 
  
When the uncertainty grows up to non-infinitesimal sizes, i.e. macroscopic 
scales,  
the perturbation $\delta$ 
is governed by the nonlinear terms and that renders its growth rate  
a scale-dependent index (Aurell et al., 1996, 1997; Artale et al., 
1997).
It is useful to introduce the Finite Scale 
Lyapunov Exponent (FSLE), $\lambda(\delta)$. Assuming $r > 1$ 
is a fixed amplification ratio and $<{\tau}_r(\delta)>$ the 
mean time that $\delta$ takes to grow up to $r \cdot \delta$, we have: 
\beq
\lambda(\delta) \,=\, {1 \over <{\tau}_r(\delta)> } \, \ln \, r
\eeq
The average $<\cdot>$ is performed over all the trajectory pairs in a 
cluster.     
We note the following 
properties of the FSLE:
\begin{itemize} 
\item[a)]{ in the limit of infinitesimal separation between trajectories, 
$\delta \to 0$, 
	the FLSE tends to the maximum Lyapunov exponent
	(MLE)};
\item[b)]{ in case of standard diffusion, $<\delta(t)^2> \sim t$, 
we find that $\lambda(\delta) \sim \delta^{-2}$ and the proportionality 
constant is of the order of the diffusion coefficient};
\item[c)]{ any slope $> -2$ for $\lambda(\delta)\; vs \; \delta$ indicates 
super-diffusive behavior, i.e. non-neglectable correlations persist 
at long times and advection is still relevant}; 
\item[d)]{ in particular, when $\lambda(\delta)=constant$ over a range 
of scales, we have exponential separation between trajectories 
at constant rate, within that range of scales (chaotic advection)}.
\end{itemize}  

Another interesting quantity related to the FSLE is the 
Lagrangian Structure Function (LSF) 
$\nu(\delta)$, defined as 
\beq
\nu(\delta) \,=\, <||{d{\bf{x'}} \over dt} - {d{\bf{x}} \over dt}||>_{\delta} 
\eeq
where the value of the velocity difference is taken at the times 
for which the 
distance between the trajectories enters the scale $\delta$ and the average 
is performed over a large number of realizations. 
The LSF, 
$\nu(\delta)$,  is a measure of the velocity at which two 
trajectories depart from each other, as a function of scale. 
By dimensional arguments, we expect that the LSF is proportional to the 
scale of the separation and to the FSLE:
\beq
\nu(\delta) \, \sim \, \delta \; \lambda(\delta)
\eeq
so that we should find similar behavior for $\lambda(\delta)$ and 
$\nu(\delta) / \delta$, if independently measured.    

In order to study the transport properties of the drifter trajectories, 
we have focused our interest on the measurement of 
$S^{2}_i(t)$, $\lambda(\delta)$ and $\nu(\delta)$. 

With regards to the practical definitions of the FSLE and the LSF, 
we have chosen a range of scales $\delta=(\delta_0, \delta_1, ..., \delta_n)$ 
separated by a factor $r>1$ such that $\delta_{i+1} = r \cdot \delta_i$ for 
$i=0,n-1$. 
The ratio $r$ is often referred to as the ``doubling'' factor  
even though it is not necessarily equal to $2$, e.g. in our case 
we fixed it at $\sqrt{2}$. 
The $r$ value has naturally an inferior bound because of the temporal 
finite resolution of the trajectories (i.e. it cannot be arbitrarily close 
to $1$) and it must be not much larger than $1$, if we want to resolve  
scale separation in the system.    

The smallest threshold, $\delta_0$, is 
placed just above the initial mean separation between two drifters, 
$\sim 10 \;  km$,  and the 
largest one, $\delta_n$, is naturally selected by the finite 
size of the domain, $\sim 500 \; km$.

Following the same procedure, 
it is straightforward to compute the LSF as the mean velocity 
difference between two trajectories at the moment in which 
the separation reaches a scale $\delta$:
\beq     
\nu(\delta) \,=\, < \sqrt{(u_1-u_2)^2 + (v_1-v_2)^2} >_{\delta}
\eeq
where the average is performed over the number of all the pairs 
within a set of particles, at the time in which 
$||{\bf{x'}}-{\bf{x}}||=\delta$. 

In section 4 below we shall show  
the results of our data analysis and compare them  
with the simulations from our chaotic model 
for the Lagrangian dynamics of the Adriatic drifters.

\section{The chaotic model}

In phenomenological kinetic modeling of geophysical flows, 
two possible approaches can be considered: stochastic and 
chaotic. Both procedures generally involve
a mean velocity field, which gives the motion over large scales, and 
a perturbation, which describes the action of 
the small scales. 
The model is stochastic or chaotic if the perturbation 
is a random process or a deterministic time-dependent function, respectively. 
Examples on kinematic mechanisms proposed to model the mixing process 
can be found in 
Bower (1991), Samelson (1992), Bower and Lozier (1994),
Cencini et al. (1999). 

The choice of one or the other depends on what 
one is interested in, and what experimental information is available. 
In our case, we have opted for a deterministic model since 
there are indications that, at the sea surface, the instabilities of the 
Eulerian structures are mostly due to air-sea interactions, 
which are nearly periodic perturbations.  

We want to consider a simple model, 
so let us assume as 
main features of the surface circulation the following 
elements: 
an anti-clockwise coastal current; two large cyclonic 
gyres; and some natural irregularities in the 
Lagrangian motion induced by the small scale structures.

Let us notice that 
the actual drifters may leave the Adriatic sea through the Otranto Strait,
but we model our domain with a closed basin, in order to study 
the effects of the finite scales on the transport, and treat it like 
a $2-D$ system, since the drifters explore the circulation in the 
upper layer of the sea, within the first meters of water. 

On the basis of the previous considerations, we introduce our 
kinematic model for the Lagrangian dynamics. 
Under the incompressibility hypothesis we write 
a $2-D$ velocity field in terms of a stream function:

\beq
u = - {\partial \Psi \over \partial y} \;\; \hbox{and} \;\; 
v = {\partial \Psi \over \partial x}.
\eeq
       
Let us write our stream function as a sum of three terms: 
\beq
\Psi(x,y,t)=\Psi_0(x,y) + \Psi_1(x,y,t) + \Psi_2(x,y,t)  
\eeq
defined as follows:
\beq
\Psi_0(x,y)={C_0 \over k_0} \cdot [- sin(k_0 (y + \pi)) +
cos(k_0 (x + 2 \pi))] 
\eeq
\beq
\Psi_1(x,y,t)={C_1 \over k_1 } \cdot sin(k_1 (x + \epsilon_1 sin(\omega_1 t))) 
sin(k_1 (y + \epsilon_1 sin(\omega_1 t + \phi_1)))               
\eeq
\beq
\Psi_2(x,y,t)={C_2 \over k_2} \cdot sin(k_2 (x + \epsilon_2 sin(\omega_2 t))) 
sin(k_2 (y + \epsilon_2 sin(\omega_2 t + \phi_2)))              
\eeq
where $k_i \,=\, 2 \pi / \lambda_i$, for $i=0,1,2$, the $\lambda_i$ 
are the wavelengths of the spatial structure of the flow; analogously 
$\omega_j \,=\, 2 \pi / T_j$, for $j=1,2$, and the $T_j$ 
are the periods 
of the perturbations. 
In the non-dimensional expression of the equations, 
the units of length and time have been set to 
$200 \; km$ and $5 \; days$, respectively. 
The choice of the values of the parameters is discussed below. 

The stationary term 
$\Psi_0$ defines the boundary large scale circulation with positive vorticity. 
$\Psi_1$ contains 
the two cyclonic gyres and it is explicitly time-dependent through a periodic 
perturbation of the streamlines. The term $\Psi_2$  
gives the motion over scales smaller than the size of the large gyres 
and it is time-dependent as well. 
A plot of the $\Psi$-isolines at fixed time is shown in Fig. 2. The actual 
basin is the inner region with negative $\Psi$ values and the zero isoline 
is taken as a dynamical barrier which defines the boundary of the domain. 

The main difference with reality is that the model domain 
is strictly a closed 
basin whereas the Adriatic Sea communicates with the rest of the Mediterranean 
through the Otranto Strait. That is not crucial 
as long as we observe the 
two evolutions, of experimental and model trajectories, within time 
scales smaller than the mean exit time 
from the sea, typically of the order of a few months.  
Furthermore, the presence of the quasi-steady cyclonic coastal current 
is compatible with the interplay between the Po river  
southward inflow at the 
north-western side and the Otranto channel northward inflow at the 
south-eastern side of the sea.  

The non-stationarity of the stream function is a necessary feature of 
a $2-D$ velocity field in order to have Lagrangian chaos 
and  mixing properties, 
that is, so that
a fluid particle will visit any portion of the domain after 
a sufficiently long interval of time. 
  
We have chosen the parameters as follows. 
The velocity scales $C_0$, $C_1$ and $C_2$ are all equal to 1, which, 
in physical dimensions, corresponds to $\sim \; 0.5 \; m/s$. 
The wave numbers $k_0$, $k_1$ and $k_2$ are fixed at $1/2$, $1$ and 
$4 \pi$, respectively. In Fig. 2a,b we can see 
two snapshots of the streamlines at fixed time.  
The length scales of the model Eulerian structures  
are of $\sim \; 1000 \; km$ (coastal current), 
$\sim \; 200 \; km$ (gyres) and 
$\sim \; 50 \; km$  (eddies). The typical 
recirculation times, for gyres and eddies, 
turn out to be of the order of 1 month and a few days, respectively. 
   
As regards to the time-dependent terms in the stream function, 
the pulsations are $\omega_1 \,=\, 1$ and $\omega_2 \,=\, 2 \pi$, 
which determine    
oscillations of the two large-scale vortices   
over a period $T_1 \simeq 30 \; days $ and oscillations of the 
small-scale vortices over a period $T_2 \simeq 5 \; days$; 
the respective oscillation amplitudes are   
$\epsilon_1 = \pi/5$ and $\epsilon_2 = \epsilon_1/10$ which   
correspond to $\sim 100 \; km$ and 
$\sim 10 \; km$.  

The choice of the phase factors, $\phi_1$ and $\phi_2$, 
determines how much the vortex pattern changes during a perturbation period. 
We have chosen to set both $\phi_1$ and $\phi_2$ to $\pi/4 \; rad$. 
  
This choice of the parameters for 
the time-dependent  
terms in the stream function is only supposed to be physically reasonable, 
for the experimental data give us limited information about 
the time 
variability of the Eulerian structures.   

The chaotic advection (Ottino, 1989; 
Crisanti et al., 1991), occurring in our model, makes an ensemble 
of initially close trajectories spread apart from one another, until 
the size of the mean relative displacement reaches a 
saturation value corresponding to the finite length scale of the domain.
  
The scale-dependent degree of chaos is given by the FSLE. 
Because of the relatively 
sharp separation between large and small scales in the model, 
we expect $\lambda(\delta)$ to display a step-like behavior with 
two plateaus, one for each characteristic time, and 
a cut-off at scales 
comparable with the size of the domain. In the limit of small 
perturbations, the FSLE gives an estimate of the MLE of the system. 

The LSF, $\nu(\delta)$, on the other hand, 
is expected to be proportional to the size of the perturbation and to 
$\lambda$,   
as discussed in the introduction. 
Therefore the quantity $\nu(\delta)/\delta$ is 
expected to be qualitatively proportional to $\lambda(\delta)$, in the sense 
that the mean slopes have to be compatible with each other.  

In the following section we will show results of 
our simulations together with 
the outcome of the data analysis.

\section{Comparison between data and model}

The statistical quantities relative to the drifter trajectories 
have been computed according to the following prescription. 
The number of selected drifters for the analysis is 37, 
distributed in 5 different deployments in the Strait of Otranto, 
containing, respectively, 4, 9, 7, 7 and 10 drifters. 
These are the only drifter trajectories out of the whole data set 
which are long enough to study the Lagrangian motion on basin scale.    
To get as high statistics as possible, at the price of losing 
information on the seasonal variability, 
the times of all of the 37 
drifters are measured as $t-t_0$, where $t_0$ is the time
of deployment.
 Moreover, to restrict the analysis only to the Adriatic basin, we 
impose the condition that a drifter is discarded as soon as its latitude 
goes south of $39.5$ N or its longitude exceeds $19.5$ E.  
  Let us consider the reference frame in which the axes are 
aligned, respectively, with 
the short side, orthogonal to the coasts, which we call 
the transverse direction, and the long 
side, along the coasts, which we call 
the longitudinal direction.

Before the presentation of the data analysis, let us briefly discuss 
the problem of finding characteristic Lagrangian times. 
A first obvious candidate is 
\beq
\tau_L^{(1)} \,=\, {1 \over \lambda}
\eeq
Of course $\tau_L^{(1)}$ is related to small scale properties. 
Another characteristic time, at least if the diffusion is standard, is 
the so-called integral time scale (Taylor, 1921) 
\beq
\tau_L^{(2)}\,=\,{1 \over <v^2>} \int_{0}^{\infty} C(\tau)d \tau
\eeq
where $ C(\tau)=\sum_{i=1}^d <v_i(t)v_i(t+\tau)>$ is the 
Lagrangian velocity correlation function and $<v^2>$ is the velocity 
variance. We want to stress that it is always possible 
(at least in principle) 
to define $\tau_L^{(1)}$ while to compute $\tau_L^{(2)}$ 
(the integral time scale) it is 
necessary to be in a standard diffusion case (Taylor, 1921).

The relative dispersion curves along the two natural directions 
of the basin, for data and model trajectories, are shown 
in Figs. 3a and 3b. 
The curves from the numerical simulation of the model are computed 
observing the spreading of a cluster of $10^4$ initial conditions. When 
a particle reaches the boundary ($\Psi = 0$) it is eliminated. 
Along with observational and simulation data,
 we plot also a straight line 
corresponding to a standard diffusion with coefficient $10^3 \; m^2/s$,
(Falco et al., 2000). We discuss this comparison below.  
Considering the effective diffusion properties, 
one should expect that the shape of $S^{2}_{i}(t)$, before 
the saturation regime, can still be affected 
by the action of the coherent structures.   
Actually, neither the data nor the model dispersion curves display 
a clear power-law behavior, and are indeed quite irregular.  
The growth of the mean square radius of a cluster of drifters appears 
still strongly affected from the details of the system, and the saturation 
begins no later than $\sim$ 1 month ($\sim$ the largest characteristic 
Lagrangian time). This prevents any attempt at defining a diffusion 
coefficient for the effective dispersion in this system.
Although the saturation values are very similar, we can see that,  
in the intermediate range, 
the agreement between observation and simulation  
is not good.  
We point out that the trouble in reproducing the drifter 
dispersion in time does not depend much on the statistics; no matter if 
37 (data) or $10^4$ (model) trajectories, the problem is that 
the classic relative dispersion  is not 
the most suitable quantity to be measured (irregular behavior even 
at high statistics).  

Let us now discuss the FSLE results. 
The curve measured from the data has been averaged over the total number 
of pairs out of 37 trajectories ($\sim 700$), under the condition that 
the evolution of the distance between two drifters is no longer followed 
when any of the two exits the Adriatic basin (see above). 
In  Fig. 4 the FSLE's for data and model are plotted. Phenomenologically, 
fluid particle motion is 
expected to be faster at small scales and slower at large scales.   
The decrease of $\lambda(\delta)$ at increasing $\delta$  
reflects  
the presence of several scales of motions (at least two) 
involved in the dynamics. 
In particular, looking at the values of  
$\lambda(\delta)^{-1}$ at the extreme 
points of the $\delta$-range, we see that 
small-scale (mesoscale) 
dispersion has a characteristic time $\sim 4$ days and the
large-scale (gyre scale) 
dispersion has a characteristic time $\sim 1$ month. The ratio between 
 gyre scale and mesoscale is of the same order as the ratio between 
the inverse of their respective characteristic times ($\sim 10$), so 
the slope of $\lambda(\delta)$ at intermediate scales is about
$-1$. 
The fact that the slope is larger than $-2$ indicates that relative dispersion 
is faster than standard diffusion up to sub-basin scales, i.e. Lagrangian 
correlations are non-vanishing because of coherent structures.  
It is interesting to compare this Lagrangian technique of measuring 
the effective Lagrangian dispersion on finite scales to the more traditional 
technique of extracting a (standard) diffusivity parameter from 
the reconstruction of the small-scale anomalies in the velocity field 
(Falco et al., 2000). Estimates of the zonal and meridional diffusivity 
in Falco et al. (2000) are of the order of $\sim 10^3 \; m^2/s$ and 
are compatible with the value of the effective finite-scale 
diffusive coefficient given 
by the FSLE, defined as $\lambda(\delta) \cdot \delta^2$, computed at 
$\delta= 20 \; km$ ($\sim$ the mesoscale).

The FSLE computed in the numerical simulations shows  
two plateaus, one at small scales and the other at large scales, 
describing a system 
with two characteristic time scales, and presents the same behavior, 
both qualitative and quantitative,  
 as the FSLE computed for the drifter trajectories. 

It is worth noting that it is much simpler 
for the model to reproduce, even quantitatively, the
relation between characteristic times and scales of the drifter 
dynamics (FSLE) rather than the behavior of the relative dispersion in time.

The LSV, in Fig. 5, shows that the behavior of $\nu(\delta)$, 
the mean velocity difference between two particle trajectories at 
varying of the scale 
of the separation, 
is compatible with the behavior of the FSLE as expected by 
dimensional arguments, i.e. $\nu(\delta)/\delta$, 
the LSF divided by the scale at which it is 
computed, has the same slope as $\lambda(\delta)$. 
This accounts for the robustness of the information given by the 
finite-scale analysis. 
 
We see the theoretical predictions of FSLE and LSV are fairly well comparable 
with the corresponding quantities observed from the data, if we consider 
the relatively simple model which we used for the numerical simulations. 
Of course, an agreement exists because of the appropriate choice of 
the parameters of the model, capable to reproduce the correct relation 
between scales of motion and characteristic times, and 
because large-scale ($\sim$ sub-basin scales) Lagrangian dispersion is  
weakly dependent on the small-scale ($\sim$ mesoscale) 
details of the velocity field.

\section{Discussion and conclusions }

In this paper we have analyzed an experimental data set 
recorded from Lagrangian surface drifters deployed 
in the Adriatic sea.  
The data span the period from December 1994 to 
March 1996, during which five sets of drifters were released 
at different times in the vicinity of the same point on 
the eastern side of the Otranto Strait. 
Adopting a technique borrowed from the 
theory of dynamical systems, 
we studied the Lagrangian transport properties 
by measuring relative dispersions, 
$S^{2}_{i}$, finite-scale Lyapunov exponents $\lambda(\delta)$, and 
Lagrangian structure function $\nu(\delta)$. 
Relative dispersion as function of time does not provide much information, 
but an idea of the size of the domain 
where saturation sets in
at long times. The behavior of $S^{2}_{i}$ 
looks quite irregular and this is due not to poor statistics but rather 
to intrinsic reasons. In contrast, the results obtained with the FSLE, 
i.e. dispersion rates at different scales of motion,  
give a more useful description of the properties of the drifter spreading. 
In particular, $\lambda(\delta)$ detects the characteristic times associated  
to the Eulerian characteristic lengths of the system. 
 
We have also introduced 
a simple chaotic model of the Lagrangian evolution and compared it 
with the observations. 
In our point of view, the actual meaning of 
the chaotic model, in relation to 
the behavior of the drifters, is not that of a best-fitting 
model. We do not claim that the quite difficult task of modeling the marine 
surface circulation driven by wind forcing can be exploited by a simple 
dynamical system. But a simple dynamical system can give satisfactory 
results if we are interested in large-scale properties of Lagrangian 
dispersion, since they depend much on the topology of the velocity field 
and only weakly on the small-scale details of the velocity structures.  
In this respect, chaotic advection, very likely present in every geophysical 
fluid flow, 
is crucial for what concerns tracer dispersion,
since it can easily overwhelm
the effects of small-scale turbulent motions on large-scale transport 
(Crisanti et al., 1991). 
Even when a standard diffusivity parameter can be computed 
from the variance and the self-correlation time of the Lagrangian 
 velocity, its relevance for reproducing the effective dispersion 
on finite scales, in presence of coherent structures, is questionable.     
In addition, practical difficulties 
arising from both finite resolution and boundary effects suggest a revision 
of the analysis techniques to be used for studying Lagrangian motion 
on finite scales, i.e. in non-asymptotic conditions. 
Considering that, generally, there is more physical 
information in a scale-dependent indicator ($\lambda(\delta)$) 
rather than in a time function ($S^2(t)$), we come to the conclusion 
that the FSLE is a more  
appropriate tool of investigation of finite-scale transport properties.     
It is important to remark 
(Aurell et al, 1996; Artale et al., 1997; Boffetta et al., 2000) 
that, in realistic cases, 
 $\lambda(\delta)$ is not just another way to look at $S^2(t) \; 
vs \; t$, in particular it is not true that $\lambda(\delta)$ behaves 
like $(d \ln S^2(t) / dt)_{S^2=\delta^2}$. This is because $\lambda(\delta)$ 
is a quantity which characterizes Lagrangian properties at the scale 
$\delta$ in a non ambiguous way. On the contrary, $S^2(t)$ can depend 
strongly on $S^2(0)$, so that, in non-asymptotic conditions, 
it is relatively easy to get erroneous conclusions only looking at the shape 
of $S^2(t)$.  
In fact,  at a given time, 
relative dispersion inside a sub-cluster of drifters 
can be rather different from other sub-clusters, e.g. because of 
fluctuations in the cross-over time between exponential and 
diffusive regimes.  
Therefore, when performing an average
over the whole set of trajectories, 
one may obtain a quite spurious and inconclusive behavior. 
On the other hand, we have seen that 
the analysis in terms of FSLE (and LSF), studying 
the transport properties at a given spatial scale, rather than at 
a given time,  
can provide more reliable  
information on the relative dispersion of tracers.

\section{Acknowledgements}

The drifter data set used in this work was
kindly made available to us by P.-M. Poulain. 
We warmly thank  
J. Nycander, P.-M. Poulain, R. Santoleri and E. Zambianchi  
 for constructive readings of the manuscript  
and for clarifying discussions about oceanographic matters.  
We also thank E. Bohm, G. Boffetta, A. Celani, M. Cencini, 
K. D\"o\"os, D. Faggioli, D. Fanelli, 
S. Ghirlanda, D. Iudicone, A. Kozlov,  
E. Lindborg, S. Marullo, 
P. Muratore-Ginanneschi and V. Rupolo for useful discussions. 

This work was supported by a European Science Foundation 
``TAO exchange grant'' (G.L.),
by the Swedish Natural Science 
Research Council 
under contract M-AA/FU/MA 01778-334 (E.A.) and 
the Swedish Technical Research Council under contract 97-855 (E.A.),
and by the  I.N.F.M. ``Progetto 
di Ricerca Avanzata TURBO'' (A.V.) and MURST, program 9702265437 (A.V.).  
G.L. thanks the K.T.H. (Royal Institute of Technology) 
in Stockholm for hospitality. 
We thank the European Science Foundation and the
organizers of the 1999 Tao Study Center for
invitations, and for an opportunity
to write up this work.


\newpage

\centerline{\bf REFERENCES}

\begin{description}

\item { Adler R.J., P. M\"uller and B. Rozovskii (eds.). 1996. 
Stochastic modeling in physical oceanography. Birkh\"auser, Boston.}

\item { Artale V., G. Boffetta, A. Celani , M. Cencini and
A. Vulpiani. 1997. Dispersion of passive tracers in closed basins:
beyond the diffusion coefficient. Phys. 
of Fluids, 9, 3162.}

\item { Artegiani A., D. Bregant, E. Paschini, N. Pinardi, 
F. Raicich and A. Russo. 1997. 
The Adriatic Sea general circulation, parts I and II. 
J. Phys. Oceanogr., 27, 8, 1492-1532.} 

\item{ Aurell E., Boffetta G., Crisanti A., Paladin G., Vulpiani A. 1996.
Predictability in systems with
many degrees of freedom.
Physical Review E, 53, 2337.}

\item{
Aurell, E., G. Boffetta, A. Crisanti, G. Paladin and A. Vulpiani. 1996. 
Growth of non-infinitesimal perturbations in turbulence.
Phys. Rev. Lett., 77, 1262-1265.}  

\item{
Aurell, E., G. Boffetta, A. Crisanti, G. Paladin and A. Vulpiani. 1997. 
Predictability in the large: an extension of the concept of Lyapunov 
exponent. J. of Phys. A, 30, 1.}

\item { Benettin G., L. Galgani, A. Giorgilli and 
J.M. Strelcyn. 1980. Lyapunov characteristic exponents for
smooth dynamical systems and for Hamiltonian systems:
a method for computing all of them. 
 Meccanica, 15, 9.}

\item { Boffetta G., A. Celani, M. Cencini, G. Lacorata and A. Vulpiani. 
2000. 
Non-asymptotic properties of transport and mixing. 
Chaos, vol. 10, {\it 1}, 50-60, 2000.}

\item {Boffetta G., M. Cencini, S. Espa and G. Querzoli. 1999. 
Experimental evidence of chaotic advection in a convective
flow. Europhys. Lett. 48, 629-633.}

\item {Bouchaud J.P. and A. Georges. 1990. 
Anomalous diffusion in disordered media: statistical mechanics, models 
and physical applications. Phys. Rep., 195, 127.}

\item {Bower A.S. 1991. A simple kinematic mechanism 
for mixing fluid parcels across a meandering jet.
J. Phys. Oceanogr., 21, 173.}

\item {Bower A.S. and M.S. Lozier, 1994. 
A closer look at particle exchange in the Gulf Stream. 
J. Phys. Oceanogr., 24, 1399.}

\item {Buffoni G., P. Falco, A. Griffa and
E. Zambianchi. 1997. Dispersion processes and residence times in 
a semi-enclosed basin with recirculating gyres. The case of 
Tirrenian Sea. J. Geophys. Res., 102, C8, 18699.}

\item {Cencini M., G. Lacorata, A. Vulpiani, E. Zambianchi. 1999. 
Mixing in a meandering jet: a Markovian approach.  
J. Phys. Oceanogr. 29, 2578-2594.}

\item {Crisanti A., M. Falcioni, G. Paladin and A. Vulpiani.
1991. Lagrangian Chaos: Transport, Mixing and Diffusion in Fluids.
 La Rivista del Nuovo Cimento, 14, 1.}

\item {Davis E.E. 1985. Drifter observation of coastal 
currents during CODE. The method and descriptive view. 
 J. Geophys. Res., 90, 4741-4755.}

\item {Falco P., A. Griffa, P.-M. Poulain and E. Zambianchi. 2000. 
Transport properties
in the Adriatic Sea as deduced from drifter data.
J. Phys. Oceanogr., in press.}
  
\item {Figueroa H.O. and D.B. Olson. 1994.
Eddy resolution versus eddy diffusion in a double gyre GCM. 
Part I: the Lagrangian and Eulerian description.
J. Phys. Oceanogr., 24, 371-386.}

\item {Hansen D.V. and P.-M. Poulain. 1996. 
Processing of WOCE/TOGA drifter data. 
J. Atmos. Oceanic Technol., 13, 900-909.}

\item {Lichtenberg A.J. and M.A. Lieberman. 
1992. Regular and Chaotic Dynamics, Springer-Verlag.}

\item {Orlic M., M. Gacic and P.E. La Violette. 1992.
The currents and circulation of the Adriatic Sea. 
Oceanol. Acta, 15, 109-124. }

\item {Ottino J.M. 1989. The kinematic of mixing: 
stretching, chaos and transport. Cambridge University.}

\item{Poulain, P.-M., A. Warn-Varnas and P.P. Niiler. 1996.
Near-surface circulation of the Nordic seas as
measured by Lagrangian drifters. 
J. Geophys. Res., 101(C8), 18237-18258.}

\item {Poulain P.-M. 1999. Drifter observations 
of surface circulation in the Adriatic sea between December 1994 and 
March 1996.  J. Mar. Sys., 20, 231-253.}

\item {Poulain P.-M. and P. Zanasca. 1998.
 Drifter observation in the Adriatic sea (1994-1996). 
 Data report, SACLANTCEN Memorandum, SM, SACLANT Undersea Research Centre, 
 La Spezia, Italy. In press.}
 

\item {Samelson R.M. 1992. Fluid exchange across a meandering Jet.
J. Phys. Oceanogr.,  22, 431.}

\item {Samelson R.M. 1996. Chaotic transport by mesoscale motions, 
in R.J. Adler, P. Muller, B.L. Rozovskii (eds.):  Stochastic 
Modeling in Physical Oceanography. Birkh\"auser, Boston, 423.}
  
\item {Taylor G.I. 1921. 
Diffusion by continuous movements.
 Proc. Lond. Math Soc. (2) 20, 196-212.}

\item {Yang H., 1996. Chaotic transport 
and mixing by ocean gyre circulation, in R.J. Adler, P. Muller, 
B.L. Rozovskii (eds.):  Stochastic Modeling in Physical Oceanography,
Birkh\"auser, Boston, 439.}

\item {Zore M. 1956. 
On gradient currents in the Adriatic Sea. 
Acta Adriatic, 8 (6), 1-38.} 

\end{description}

\newpage
\centerline{\bf FIGURE CAPTIONS}

\begin{description}

\item{FIGURE 1: Plot of the 37 drifter trajectories in the Adriatic Sea 
used for the data analysis. The 
longitude east and latitude north coordinates are in degrees. The drifters 
were deployed in the eastern side of the Otranto strait. }
\item{FIGURE 2: Model stream function isolines at a) $t=0$ and 
b) $t=T_1/2$, with   
$T_1 \sim 30$ days. The 
boundary of the domain is the zero isoline. The  
coordinates ($ x, \; y$) are in $km$.}
\item{FIGURE 3: Relative dispersion curves, 
for data (plus symbol) 
and model (dashed) trajectories, along the two natural 
directions in the basin geometry: 
a) transverse component, b) longitudinal component. The time 
is measured in $days$ and the dispersion in $km^2$. 
The experimental curve is computed over the 37 drifter trajectories; 
the model curve is computed over a cluster of $10^4$ particles, 
initially placed at the border of the southern gyre with a mean 
square displacement of $\sim \; 50 \; km^2$. 
The straight line with slope 1 has been plotted for comparison 
with a standard diffusive scaling with a corresponding diffusion 
coefficient $\sim 10^3 \; m^2/s$, typical of marine turbulent motions. 
}   
\item{FIGURE 4: Finite-Scale Lyapunov Exponent for data (continuous line) 
and model (dashed line) trajectories. $\delta$ is in $km$ and 
$\lambda(\delta)$ is in $day^{-1}$. 
The experimental FSLE is computed over all the pairs of trajectories    
out of 37 drifters; the FSLE from the model is averaged over $10^4$ 
simulations. 
The simulated $\lambda(\delta)$  
 has a step-like behavior with one 
plateau at small scales ($< \, 50 \; km$) and one at basin scales 
($> \, 100 \; km$), 
corresponding to doubling times of $\sim 3 \; days$ and 
$\sim 30 \; days$, respectively. }
\item{FIGURE 5: Lagrangian Structure Function for data (continuous line) 
and model (dashed line) trajectories. $\delta$ is in $km$ and 
$\nu(\delta)/\delta$ is in $day^{-1}$. The experimental LSF is computed 
over all the pairs of drifters; the LSF from the model 
is averaged over $ 10^4$ 
simulations.}

\end{description}


%






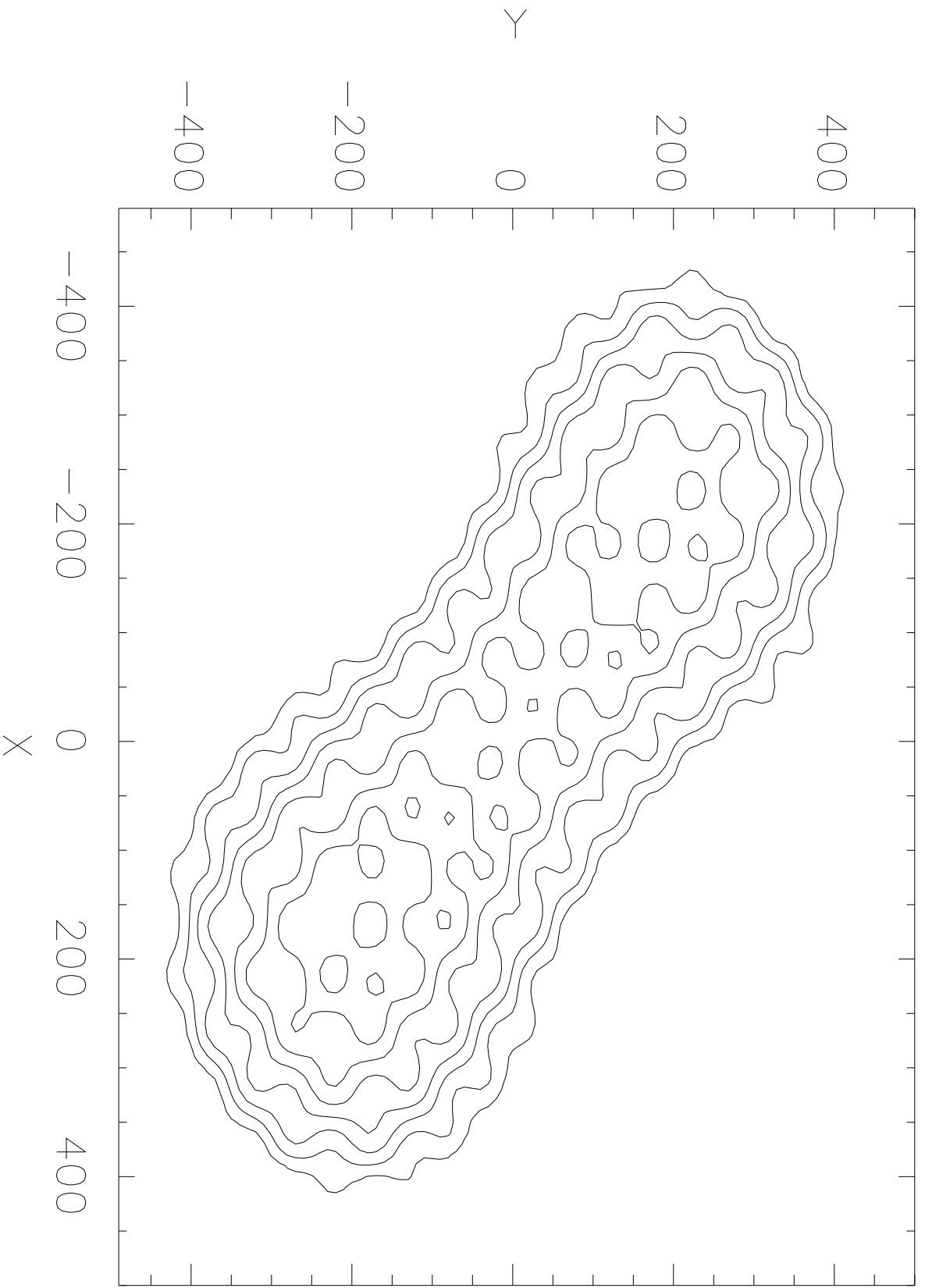
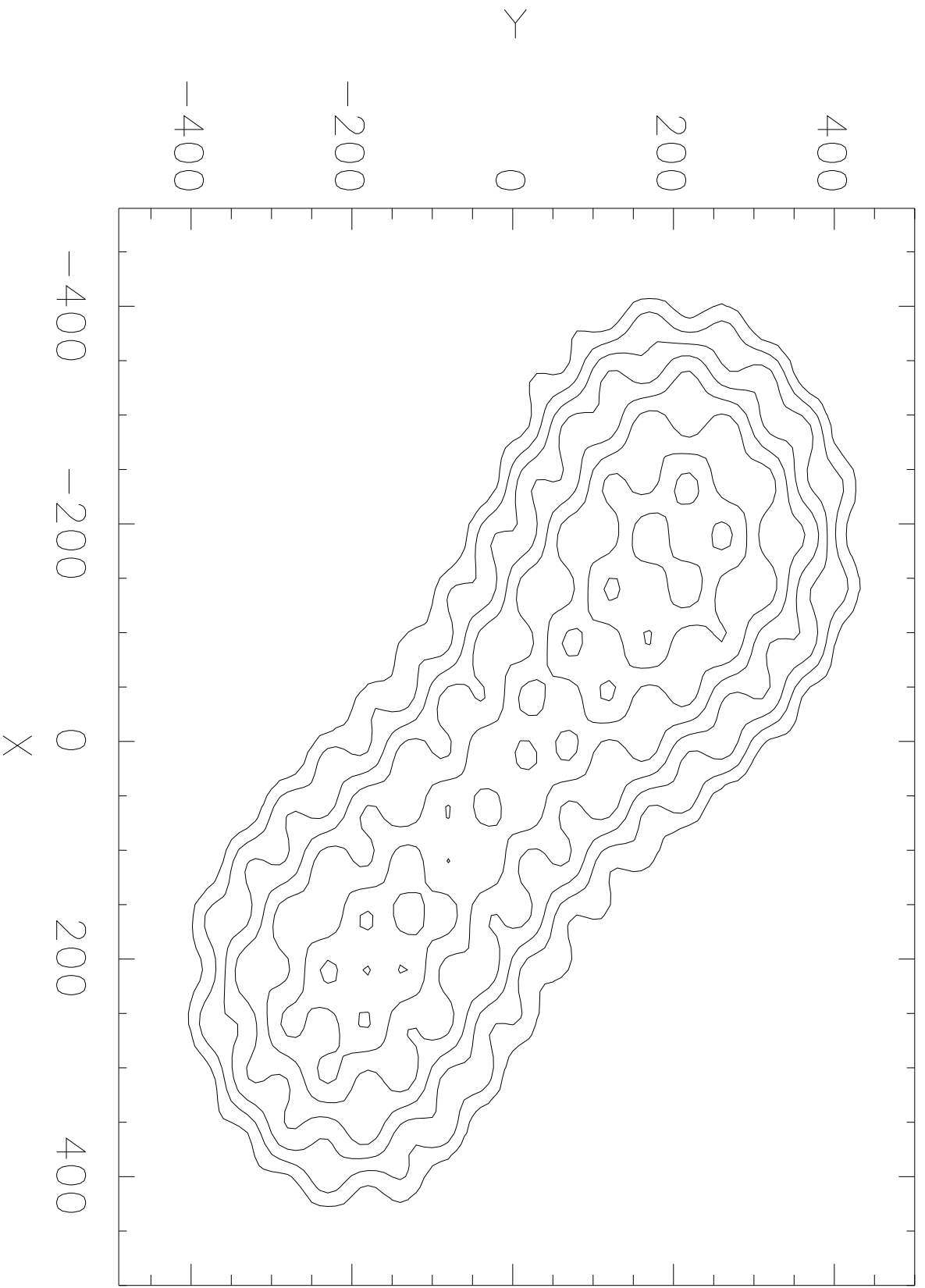
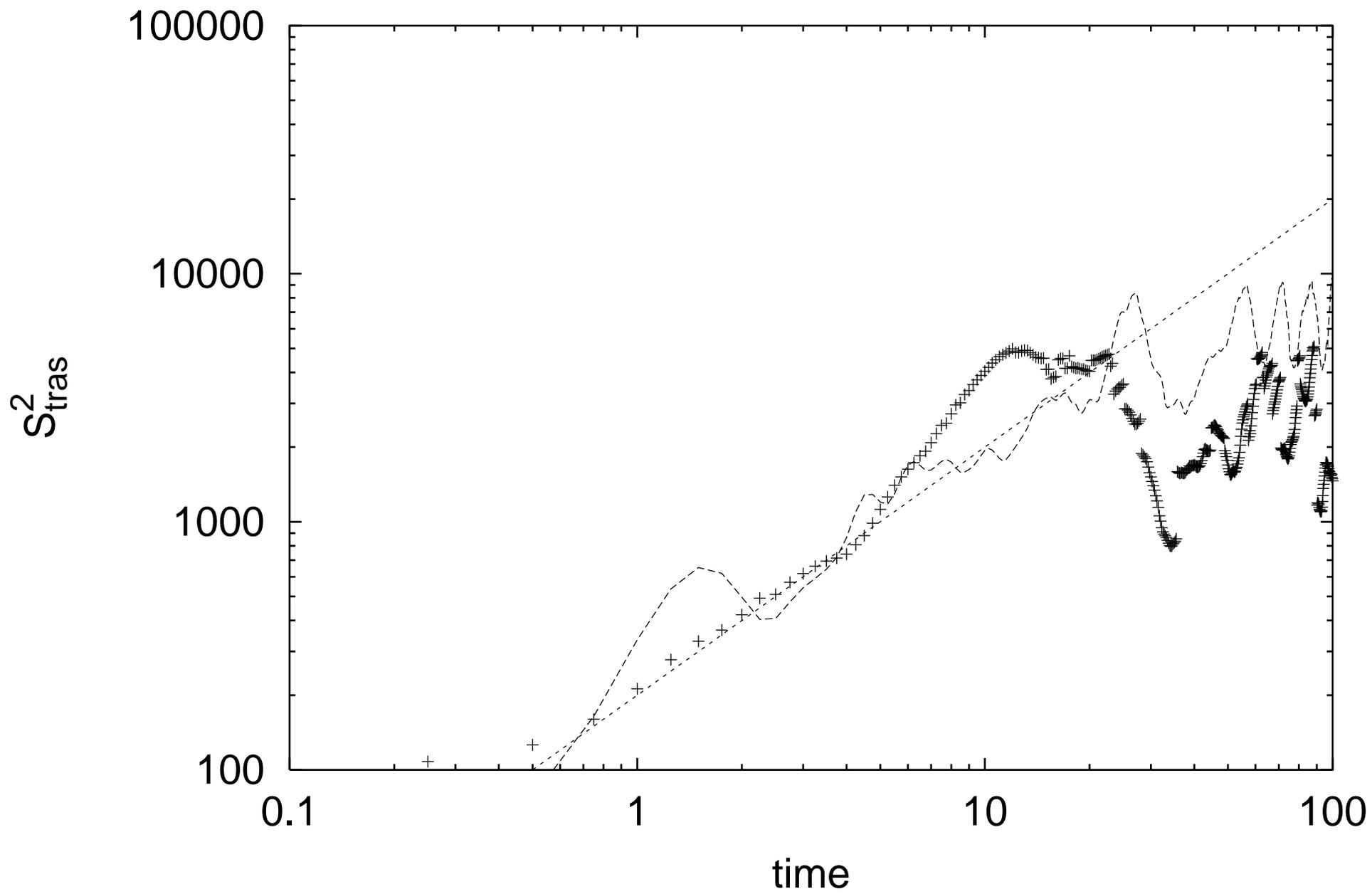
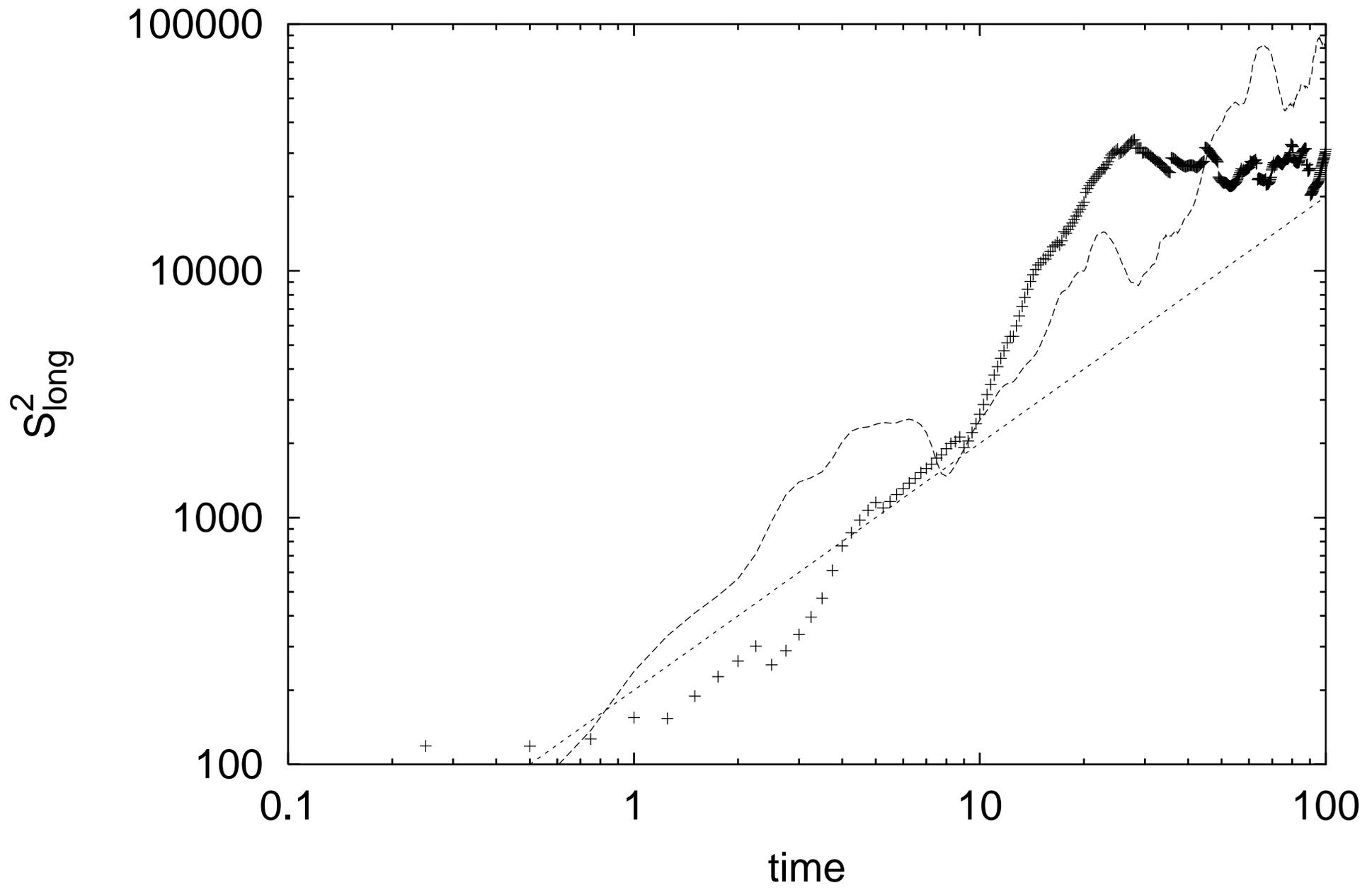
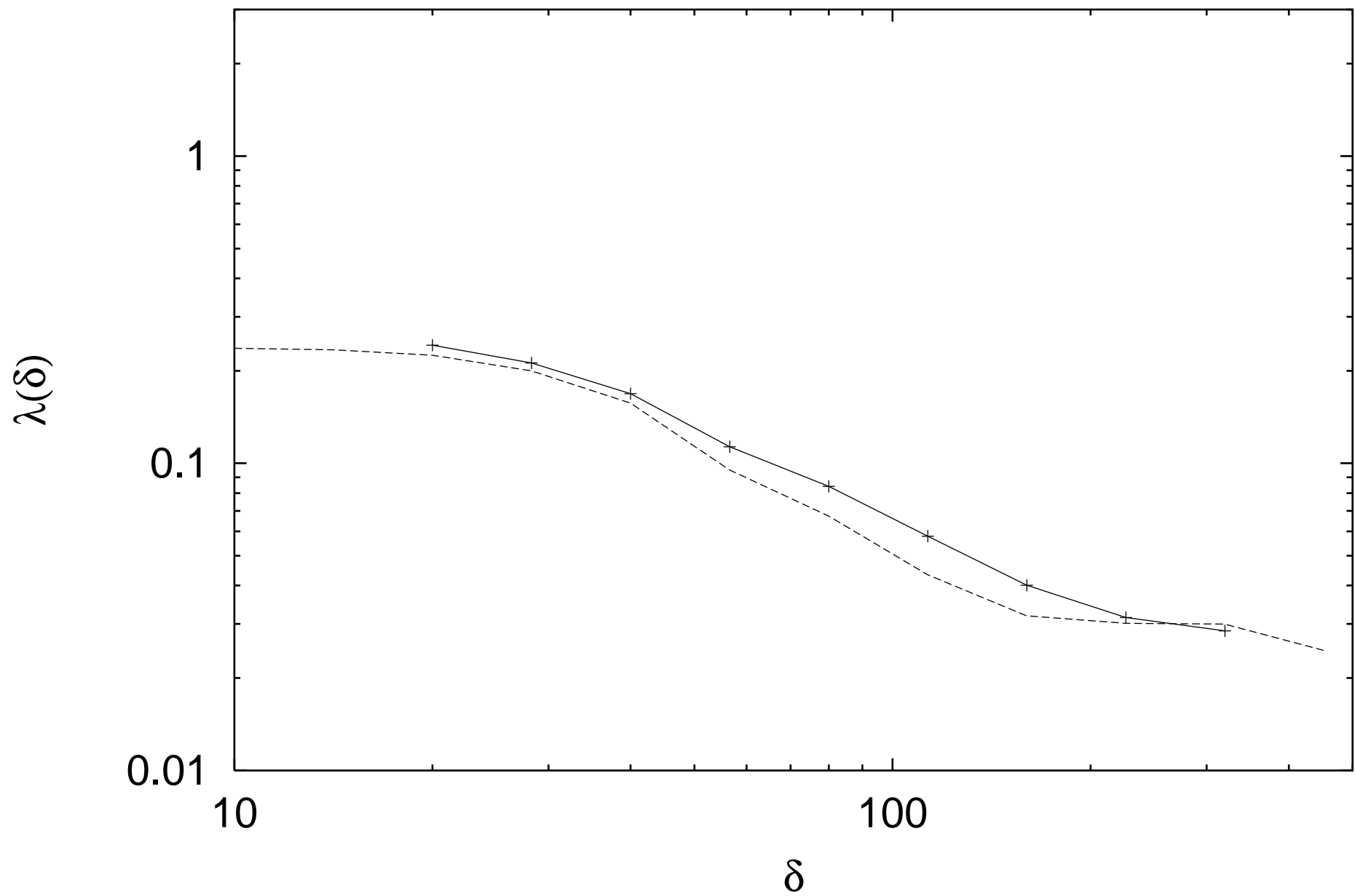
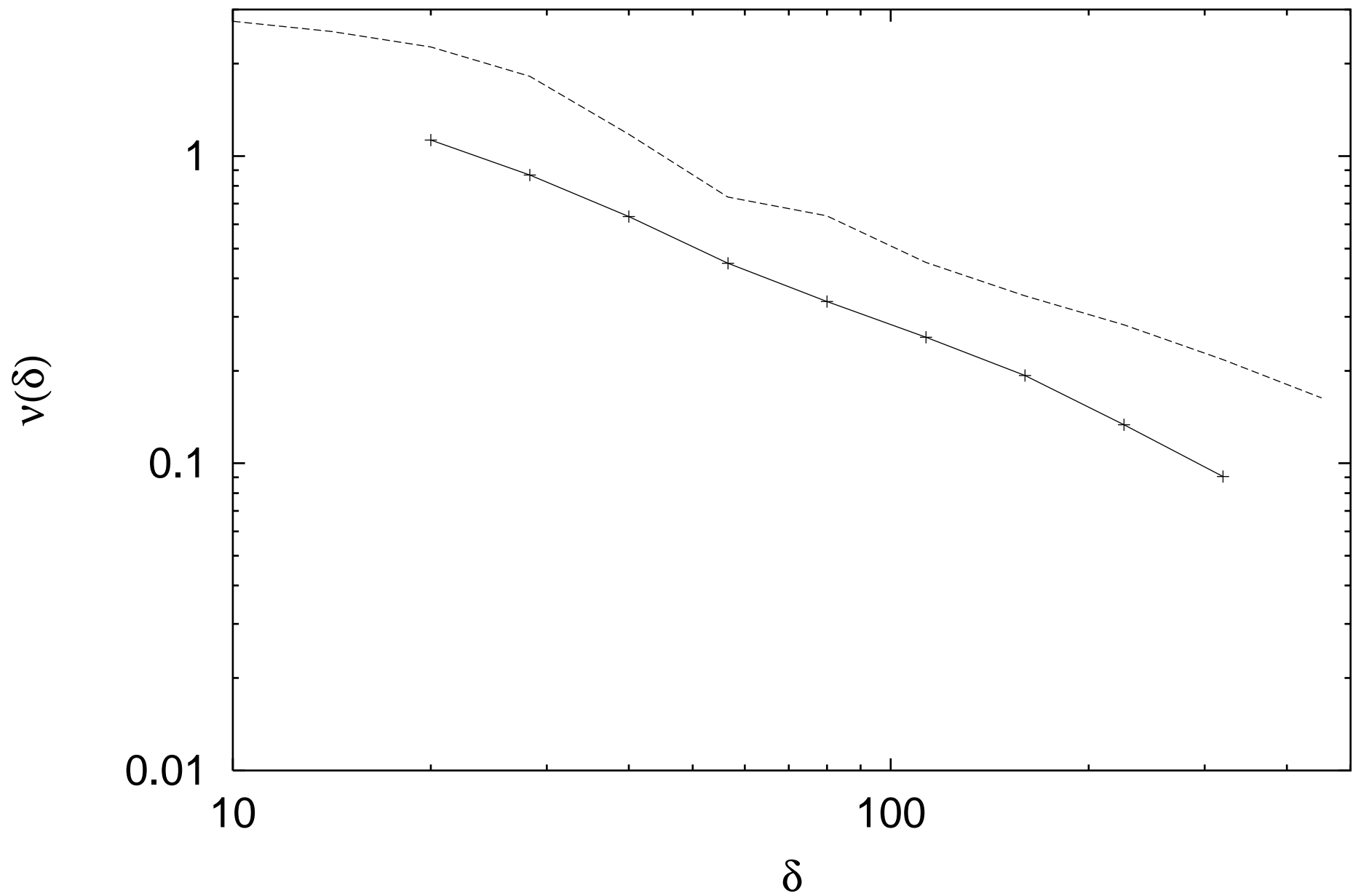
\end{document}